\documentclass[graybox]{svmult}
\usepackage{mathptmx}       
\usepackage{helvet}         
\usepackage{courier}        
\usepackage{type1cm}        
\usepackage{makeidx}         
\usepackage{graphicx}        
\usepackage{multicol}        
\usepackage[bottom]{footmisc}
\begin{document}

\title{Quantum Machine Learning}
\author{Muhammad Usman}

\institute{{{Professor Muhammad Usman} \at{Data61, CSIRO, research Way Clayton 3168 Australia and School of Physics, The University of Melbourne, Parkville 3010 Victoria Australia and School of Physics, Monash University, Clayton 3166 Victoria Australia}, \email{muhammad.usman@data61.csiro.au}}}

\maketitle
\label{label}


\begin{abstract}\\
The meteoric rise of artificial intelligence in recent years has seen machine learning methods become ubiquitous in modern science, technology, and industry. Concurrently, the emergence of programmable quantum computers, coupled with the expectation that large-scale fault-tolerant machines will follow in the near to medium-term future, has led to much speculation about the prospect of quantum machine learning (QML), namely machine learning (ML) solutions which take advantage of quantum properties to outperform their classical counterparts. Indeed, QML is widely considered as one of the front-running use cases for quantum computing. In recent years, research in QML has gained significant global momentum. In this chapter, we introduce the fundamentals of QML and provide a brief overview of the recent progress and future trends in the field of QML. We highlight key opportunities for achieving quantum advantage in ML tasks, as well as describe some open challenges currently facing the field of QML. Specifically in the context of cybersecurity, we introduce the potential for QML in defence and security-sensitive applications, where it has been predicted that the seamless integration of quantum computing into ML will herald the development of robust and reliable QML systems, resilient against sophisticated threats arising from data manipulation and poisoning.  
 
\end{abstract}

\section{Introduction}

Quantum computing is an emerging field of research which promises immense computational power to solve challenging computational problems which are otherwise intractable by classical computing methods. Recently, researchers from Google have shown that their quantum processor could perform a computational task in less than five minutes which a conventional classical computer would take septillion years! \cite{Google_septillion}. Although a remarkable result, their work showed supremacy of the quantum processor only for a contrived task, and a general-purpose quantum advantage for real-world applications will still require significant more development on both quantum hardware and software fronts. Nevertheless, the anticipated revolutionary impact of quantum computing has led to intense research on identifying its applications in many fields of research and technologies including quantum chemistry \cite{doi:10.1021/acs.chemrev.8b00803}, drug discovery \cite{Drug_discovery}, financial optimisation \cite{Finance}, and transport \cite{Transport}. The key aim for quantum computing research is to develop fundamentally new algorithms and computing methods which can outperform conventional classical approaches. Among those, integrating the power of quantum computing in machine learning (ML) is quite intrinsic as the associate computational power from quantum computing is anticipated to lead to significant speed-up, enhanced accuracy, and/or superior robustness of ML models \cite{QML}. This has led to the birth a new field of research namely quantum machine learning (QML) -- the development and bench-marking of ML models which explicitly rely on the unique properties of quantum mechanics such as superposition and entanglement to outperform their classical counterparts \cite{QML_review, QAML_review, riste2017demonstration, liu2021rigorous}. The field of QML is currently one of the most rapidly advancing areas of research, with quantum versions of almost all classical ML algorithms being actively developed and benchmarked on the existing near-term noisy quantum processors.  

An important line of research within the field of QML is pertaining its application in security-sensitive applications such as intelligence, security, surveillance, reconnaissance, and targeting systems \cite{QAML_review}. Despite high efficiency and accuracy of classical ML algorithms, it has been found that they can be readily fooled by an adversary through manipulation or spoofing of data (also known as adversarial attacks) \cite{kurakin2016adversarial}, which poses serious security threat for applications where reliability is the key parameter of interest. This has raised an important question: whether QML algorithms are also as vulnerable as classical ML models \cite{benchmarking, lu2020quantum, PhysRevA.101.062331}. Recent preliminary work has theoretically shown that QML algorithms are remarkably robust against adversarial attacks \cite{benchmarking}, which is attributed to the fact that classical adversarial attacks are ineffective on QML models due to fundamentally lacking quantum resources such as quantum entanglement which is the hallmark of quantum systems \cite{Neil_adversarial}. This offers a unique opportunity to leverage quantum computing, specifically its unique properties like superposition and entanglement, to develop highly resistant QML based autonomous systems, leading to a new area of research known as quantum adversarial machine learning (QAML). 

Despite significant progress in algorithmic development and benchmarking of QML and QAML models, their experimental implementation on quantum processors is still in its infancy \cite{Drastic_encoding, ren2022experimental}. Further advancements in the capabilities of quantum models towards practical scale applications in particular their experimental implementation can lead to an end-to-end QAML sovereign capability to secure future autonomous Intelligence Surveillance and Reconnaissance (ISR) systems for military and Defence purpose.

\section{Analysis}

Classical ML is not a new field of research. The progress in classical ML methods has undergone decades of development; however, it is only quite recently that the state of the art classical ML such as deep neural networks, large language models and natural language processing have found remarkable applications in nearly all fields of research and industrial workflows. A key reason for such tremendous rapid progress in ML during recent days is due to the availability of tremendous computational capabilities which have enabled efficient training and testing of ML models on very large datasets. The fundamental requirement of ML for computational power indicates that it is ideally poised to benefit from quantum computing which promises tremendous computational advantage over classical supercomputers. 

It is hypothesised that many of the ML algorithms rely on linear algebra routines such as Fast Fourier transform, matrix inversion or finding the Eigen values of a large matrix, which a quantum computer might be able to solve more efficiently than a classical computer, thereby providing a speed-up to the QML models over their classical counterparts \cite{QML}. However, in order for such benefit to be practically achieved, the loading of classical data into a quantum state has to be efficient, so that the exponential cost of the data loading step does not overcome any benefit of QML model. Indeed, recent work has focused on efficient quantum state preparation for QML models \cite{Drastic_encoding}. 

Apart from training speed-up, it is also important to explore other possibilities of quantum advantage for ML tasks such as robustness against adversarial attacks. Finally, the application of QML to quantum data is considered a promising pathway to achieve quantum advantage which circumvents the exponential cost of encoding classical data \cite{quantum_data}.         

The vulnerability of ML algorithms has been well known in the classical literature as data spoofing and manipulation attacks can be designed which can easily fool even very powerful classical ML models. This has given the birth to a new field of research known as adversarial ML \cite{AML_1, AML_2}, an area of research which deals with attacks and defence of ML models. Although there have been numerous attacks and defence methods designed for classical ML models with varying degrees of success, there is no clear resolution if a universal defence method exists which can overcome the vulnerabilities of ML models. An important finding in this context is related to the high transferability of attacks from one ML model to other ML models, \textit{i.e.}, attacks designed for one specific ML model are found to be highly effective on other independently trained ML models \cite{benchmarking}. This leads to serious security concerns for ML applications in security-sensitive applications such as in Defence and military systems, as one adversary could design attack on their models and transfer to models working in highly sensitive applications causing serious damage.

With the integration of quantum computing into ML in recent years, researchers have started to investigate this important question whether QML will suffer from similar vulnerabilities \cite{QAML_review}. Recent work has focused on the analysis of QML models, in particular with the context of transferability of attacks between classical and quantum ML architectures \cite{benchmarking}. It has been discovered that while the attacks from classical ML models do not transfer to QML models, contrarily the attacks from QML models were easily able to fool classical ML algorithms. This is an important finding which describes a dual advantage for early adapters of quantum computing technologies. On one hand, quantum computing could generate highly effective attacks easily fooling classical ML systems, while on the other hand being very robust against any adversarial attack. The study also reported that QML networks remain vulnerable against quantum attacks and more work is needed to ensure complete safety of ML system in post-quantum era where an adversary may also have access to quantum computers.   

\subsection{Definition}

A typical QML model such as quantum neural network consists of three major building blocks \cite{QAML_review}: data encoding, feature learning, and measurement outcome (See Figure 1). For classical datasets, the data encoding step loads a classical dataset into a quantum state by using an appropriate encoding scheme. The common encoding schemes are amplitude encoding \cite{larose2020robust}, phase encoding \cite{larose2020robust}, and Flexible Representation for Quantum Images (FRQI) \cite{dilip2022data}. The second building block is generally a variational quantum circuit consisting of single qubit rotation gates and two-qubit entangling gates such as control-phase gates. The classical optimisation of single qubit rotation gates during the training process allows the learning of features from the input data, whereas the two-qubit gates introduce entanglement. The final step is measurement which for any test data instance reveals the outcome of the quantum machine learning model. While the data loading and measurement steps remain the same, the implementation of the feature learning block could vary based on the architectural design of a particular quantum machine learning model.     

\begin{figure*}[ht]
 \begin{center}
 \includegraphics[width=11cm]{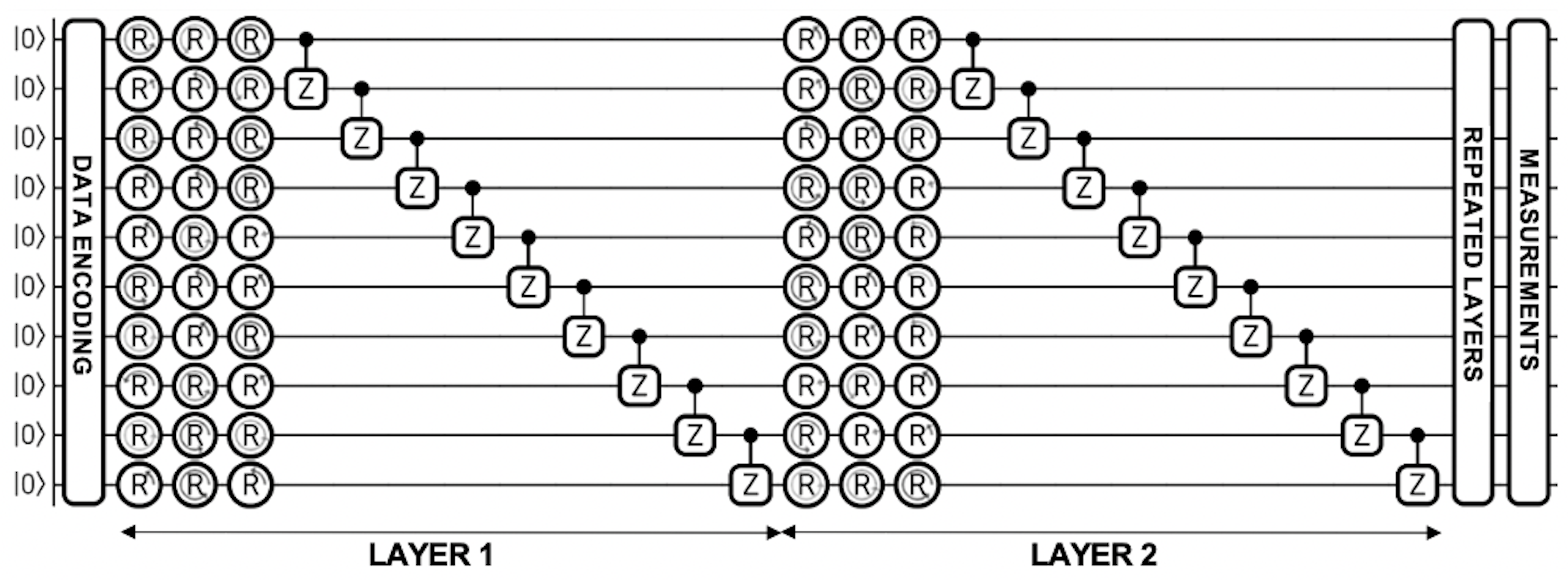}
   \caption{\label{fig:1} Schematic illustration of a standard quantum variational architecture which has been trained in the literature to perform QML tasks. The circuit consists of a data encoding layer, followed by multiple layers of trainable variational gates and two-qubit CNOTs. The final measurement layer reveals the outcome of the trained QML network.}
 \end{center}
 \end{figure*}

QAML is a field of research in which new QML models are developed and benchmarked to investigate their robustness against adversarial attacks. An adversarial attack is defined as a careful manipulation of input dataset such that it is able to flip the outcome of a well-trained ML model (quantum or classical). For example, a well-trained ML model on the photos of cats and dogs will correctly label previously unseen photos of cats and dogs. However, an effective adversarial attack will carefully manipulate a photo of cat by selectively changing some pixels in the image such that the ML model will label it incorrectly as the photo of a dog. There are two kinds of adversarial attack settings: black-box attack and white-box attack. In a black-box attack, the attack is generated without any knowledge of the ML model being attacked. This is also known as transferability of an attack, as the attack is usually generated on one model and then transferred to a different target model. Contrarily, a white-box attack is generated with complete knowledge of the ML model being attacked. Several attacks have been developed in the literature to trick ML models such as PGD \cite{pgd}, FGSM \cite{fgsm}, and Auto \cite{auto} with offer varying degrees of effectiveness.          

\subsection{Maturity}

The field of QML is very nascent at this stage, with new algorithmic developments rapidly emerging and their testing on the available quantum hardware is presently an active area of research. Despite significant progress and promising results, the application of QML models has been primarily to simple proof-of-concept datasets, whereas its implementation on datasets from real-world applications would require significant future research and new scientific breakthroughs in the coming years. There are still several key challenges associated with various components in the QML pipeline such as efficient loading of classical data into quantum states, overcoming barren plateaus in the training of QML models, mitigating or cancelling the impact of noise or errors in quantum hardware and optimisation of QML architectures to achieve accuracies at par with classical counterparts. However, recent work reported in the literature \cite{QAML_review} has shown significant progress on all fronts which incite the excitement around the prospect of QML for practical problems in the coming years.         

Likewise, on theoretical front, a few recent studies have shown that the integration of quantum computing in ML has a clear potential towards enhancing their robustness against adversarial attacks (see for example Ref. \cite{QAML_review} and references therein). The experimental implementation is still challenging due to intensive resource requirements and relatively limited capabilities of the current generation of quantum processors. Nevertheless, a couple of initial proof-of-concept experimental demonstrations of QAML models have already been reported \cite{Drastic_encoding, ren2022experimental}, confirming the potential for quantum-enhanced robustness of ML models.   

\subsection{Trends}
QML and QAML are relatively new areas of research which have gained momentum only in the last few years. Given its strong implications for security-sensitive applications, QAML has been recently a subject of strong attention in particular within the Defence and military communities. AS QML and QAML are closely related fields of research, they share many common trend. Below we highlight a few active areas of research in the context of QML and QAML development:  

\begin{enumerate}

\item \textbf{Optimisation of encoding circuits.} Data encoding is presently one of the biggest issues towards the practical implementation of QML and QAML, particularly in the era of near-term quantum devices. Two usual approaches are amplitude and phase encoding \cite{larose2020robust}, both coming with their own advantages and overheads. For example, amplitude encoding is efficient by exploiting the exponentially (in the number of qubits) large Hilbert space of the quantum computer, but its main limitation is that it requires exponentially deep encoding quantum circuits -- a significant overhead for the implementation on near-term quantum devices where noise severally limits the fidelity of deep quantum circuits. Contrarily, phase encoding, in which data is encoded into the angles of single-qubit rotations, is more efficient in regard to circuit depths, but it demands a large number of qubits, which seriously limits its ability to implement on the current quantum hardware where only a handful number of qubits are available. There are other encoding schemes such as interleaved data encoding strategy, which consists of alternating layers of data encoding and variational gates. Such a strategy allows for a user-controlled trade-off between number of qubits and circuit depth, making it particularly suitable for the current generation of quantum computers. Notably a recent experimental implementation of QAML employed this encoding scheme \cite{ren2022experimental}. Nevertheless the development of efficient data encoding schemes remains an active area of research for both QML and QAML. While the current implementation of QML is primarily focused on simple proof-of-concept datasets such as MNIST and FMNIST, its scalability towards complex real-world data sets will inevitably require new and more efficient data encoding schemes. An alternative line of research is to apply data reduction schemes by classical pre-processing or using novel methods for approximate state preparation which drastically reduce the overhead associated with data encoding \cite{Drastic_encoding, romero2017quantum}.

\item \textbf{Architecture design.} Central to the conventional approaches to QML and QAML is an optimisable variational quantum circuit sandwiched between a data encoding circuit and a set of measurements to determine the prediction of the classifier. A quantum variational circuit is made up of a repeated sequence of a number of parameterized single-qubit rotation gates followed by two-qubit gates generating entanglement. The parameters of single qubit gates are classically optimised to learn input data features, which is conceptually similar to tuning of classical neuron weights in traditional neural networks. Although variational quantum circuit approach has been quite successful for simple image datasets, its performance for larger and complex data is already facing challenges such as trainability, barren plateaus, and expressibility. As an example, the presence of barren plateaus in the training loss landscape leads to a serious impediment to the training of deep quantum circuits. Although it not entirely clear what form the large-scale quantum classifiers of the future will take, it has already started to become evident that future QAML models will require much more sophisticated architectures if they are to train on highly complex datasets relevant for practical applications. Recent work has already started which is focusing on developing quantum convolutional networks \cite{henderson2020quanvolutional} and quantum RESNET \cite{heredge} -- a quantum version of a powerful and sophisticated classical architecture. 

\item \textbf{Noise mitigation and error correction.} The current generation of quantum devices suffer from noise or errors, and therefore it is challenging to implement deep quantum circuits which are relevant for practical applications. Likewise, the experimental implementations of QML and QAML also face this crucial challenge arising from the limitations of the current quantum processors, due to limited numbers of qubits and high levels of noise. Although a proof of concept QAML study was experimentally implemented in a recent work \cite{ren2022experimental}, sophisticated QAML applications of the future will require error mitigation and correction. Another interesting line of research is to explore if the presence of noise helps in adversarial robustness of QML models \cite{du2021quantum}. It might be possible that the noise in quantum devices dilute the presence of adversarial attacks which in itself are based on the carefully crafted noise in the datasets leading to overall better performance. Significant more research is needed to fully understand both the role of noise in the working of quantum adversarial architectures as well as their performance when implemented with error correction or mitigation schemes.  

\item \textbf{Quantum generative adversarial learning.} The interplay of adversarial ML with generative adversarial networks is an interesting line of research. Generative networks have been extensively used in adversarial ML context, both in generating strong attacks and effective defence. Recently quantum generalisations of generative adversarial networks have been proposed. It consists of a generative network where either the generator or the discriminator (or both) has been implemented by a QML model \cite{Tsang}. The bench-marking of the performance of quantum adversarial generative networks in the context of either generating or detecting adversarial examples is another interesting line of research concerning hybrid quantum-classical models; for example, it could be that a quantum discriminator may exhibit superior performance in detecting adversarial perturbations which were themselves generated by a quantum network.  

\item \textbf{Quantum machine learning beyond images.} While the focus of QML and QAML has been primarily on image datasets, it is important to adapt these models for generic datasets such as signals and text to broaden the scope of their applications. A recent work has been reported on the benchmarking of QAML for signals which has shown similar robustness against adversarial attacks as previously reported for image datasets \cite{Autumn}. More work is needed to adapt QAML framework for a wide range of signal datasets including microwave, radio frequency, and radar signals and perform detailed studies to understand quantum adversarial solutions for signals.    

\item \textbf{Quantum transfer learning:} An interesting line of research is to combine QML models with classical ML techniques to create new hybrid architectures which may retain unique properties of QML while enabling applications for complex large-sized datasets. In this context, quantum transfer learning has been reported as an effective technique where a quantum variational circuit is concatenated with a classical ML model pre-trained on ImageNet dataset \cite{Amena_paper}. The resulting hybrid architecture demonstrated promising performance when benchmarked on a variety of image datasets such as CIFAR-10, traffic signs and And \& Bees. While this work was based on transfer learning from a classical to a quantum network, future research may also investigate transfer learning between two quantum architectures or from a quantum to a classical architecture. 

\item \textbf{QML exploiting data symmetries:} Optimising QML architectures by explicitly exploiting data symmetries has been a fruitful line of research in the past few years. It has been reported that the training and test accuracies can be drastically boosted by developing QML models which takes into account reflection \cite{PRXQuantum.5.030320} and rotation \cite{T_West_2023} symmetries of datasets. Although these studies have been confined to image datasets, the focus can be expanded for many other datasets as symmetries are a key characteristics of many real-world applications.    

\item \textbf{Quantum attacks.} QML offers robustness against classical adversarial attacks and therefore may lead to quantum advantage for early adapters of quantum technology. However, if an adversary also has access to a quantum computer, it is yet not fully known if QML will be robust against quantum attacks, \textit{i.e.}, attacks natively generated on quantum computers. A recent work indicated that quantum attacks are effective on simple quantum architectures based on variational classifiers \cite{benchmarking}, however a comprehensive analysis is needed to fully determine the scope and extent of attacks and defence in the context of QML.  

\item \textbf{Quantum machine learning on quantum data:} As the loading of classical data into a quantum state is an expensive step in the QML pipeline, it is considered that QML implemented directly on quantum data such as from a quantum sensing device may provide a robust pathway to achieve quantum advantage \cite{quantum_data}. However, it is not clear yet how a QML model would be trained directly on quantum data in a practical setting which will require significant research breakthrough in the future.   

\end{enumerate}

\section{Recommendation for Advancement of QML and QAML}

Quantum computing is anticipated to be a revolutionary technology which promises tremendous computational advantages over the conventional classical approaches. Its integration with ML systems is expected to transform their capabilities, with benefits including speed in training, novel feature extraction, and protection against data spoofing and cyber attacks, overcoming the vulnerabilities of the current systems deployed in military and Defence applications. Despite many theoretical studies indicating potential advantage of QML and QAML, its experimental implementation is still at its very early stages and face challenges arising from noisy hardware with limited number of qubits which limit their capability to tackle deep quantum circuits typically required for QML and QAML tasks. Our recommendations to advance QML and QAML include design of optimised architectures with reduced circuit depths, development of efficient data encoding schemes and benchmarking of QML/QAML models with error mitigation and error correction schemes. 

\section{Conclusion}

Quantum machine learning is a rapidly advancing field of research which has the potential to offer transformational socio-economic impact by addressing challenging problems such as personalised drugs for complex deceases like cancer, climate science by finding novel materials for decarbonisation, congestion control through everyday traffic modelling, or predicting future financial crisis by modelling large banking and stock datasets. The current state of the field is largely at the proof-of-concept level where new quantum machine learning models are being developed and tested on the near-term quantum devices. With remarkable advancements in the scalability and quality of quantum processors, it is anticipated that quantum machine learning will also demonstrate increasingly capable performances in the next few years and some selective applications could start emerging with possible advantage over classical methods.          

The integration of quantum computing in ML and AI systems will have important implications for the robustness of future autonomous military systems including electronic/cyber warfare and the rapid determination of best countermeasures against unknown or emerging threats. Ultimately a quantum machine learning based system will be a self-contained system emulating all aspects of the quantum adversarial ML approach for threat identification and mitigation in state-of-the-art and future command, communication and control, intelligence and surveillance systems, and autonomous robotic systems underpinned by quantum algorithms.

\section{Biography}
Professor Muhammad Usman is Head of Quantum Systems and Principal Staff Researcher at CSIRO's Data61 which is Australia National Research Organisation. He has over 15 years of research and teaching experience in the field of quantum computing, with research interests in quantum algorithms, quantum software engineering, and quantum security. He is serving on the executive editorial boards of two IOP journals (Nano Futures and MSMSE), a Fellow of the Australian Institute of Physics (AIP), and have honorary Associate Professor positions at the University of Melbourne and Monash University. Dr Usman was nominated as Innovative of the Year 2023 Award by Defence Industry, Winner of the Australian Army Quantum Technology Challenge in three years in a row (2021, 2022 and 2023), Rising Stars in Computational Materials Science by Elsevier in 2020, and Dean’s Award for Excellence in Research (Early Career) at the University of Melbourne in 2019. He served as the chair of the organising committee for 8th International Conference on Quantum Techniques in Machine Learning (QTML 2024) held in Australia.

\bibliographystyle{unsrt}
\bibliography{myChapter_ref}


\end{document}